\documentclass[a4paper]{jpconf}
\usepackage{graphicx}
\usepackage{bm}
\usepackage{amsmath}
\usepackage{accents}
\def\up{\uparrow}
\def\down{\downarrow }
\def\Vec#1{\bm{#1}}

\begin{document}
\title{Time-reversal symmetry breaking phase and gapped surface states in $d$-wave nanoscale superconductors}

\author{Yuki Nagai$^{1,2}$, Yukihiro Ota$^{3}$, and K. Tanaka$^{4}$}

\address{$^{1}$Japan Atomic Energy Agency, CCSE, Kashiwa, Japan}
\address{$^{2}$Massachusetts Institute of Technology, Department of Physics, Cambridge, MA, USA} 

\address{$^{3}$Research Organization for Information Science and Technology (RIST), Kobe, Japan}

\address{$^{4}$University of Saskatchewan, Department of Physics and Engineering Physics, Saskatoon, SK, Canada}

\ead{nagai.yuki@jaea.go.jp}

\begin{abstract}
We solve the Bogoliubov-de Gennes equations self-consistently for $d$-wave superconductors with [110] surfaces.
We find spontaneous breaking of time-reversal symmetry (TRS) caused by the spontaneous occurrence of a 
complex order parameter with an extended $s$-wave symmetry along the [110] surfaces.
In the TRS-breaking phase, the $d$-wave order parameter itself becomes complex.
We also show that vortex-antivortex pairs can appear along the [110] surfaces of nanoribbons. These pairs are detectable with surface sensitive probes.
\end{abstract}

\section{Introduction}

Time-reversal symmetry (TRS) and the corresponding topological phenomena are hot topics in condensed matter physics. 
The authors of Ref.~\cite{Satoflat} have shown that the bulk-edge correspondence with respect to a topological invariant protected by TRS underlies the zero-energy Andreev bound states on [110] surfaces of high-$T_{c}$ cuprates. 
While the Andreev bound states are usually understood as arising from the sign-changing property of the 
$d_{x^{2}-y^{2}}$-wave order parameter \cite{Tsuei,Kashiwaya}, such zero-energy surface states are robust as the bulk-edge correspondence ensures the existence of gapless surface states in a topologically nontrivial system \cite{Hasan,Alicea}. 

The gapless surface states linked to TRS can become gapped by breaking of TRS \cite{Chen,Yu}.  
Spontaneous TRS breaking in $d$-wave superconductors has been discussed phenomenologically by introducing a secondary order parameter with relative phase $\pi/2$ to the predominant component that is
stabilized near the [110] surfaces 
\cite{Matsumoto,Fogelstroem}. 
This has been a controversial topic in the past two decades owing to contradicting experimental results on high-$T_{\rm c}$ cuprates \cite{Gustafsson,Hakansson}. Recent experiment, however, has clearly detected a full gap in the excitation spectrum that is consistent with spontaneously broken TRS in a nanoscale YBa$_{2}$Cu$_{3}$O$_{7-\delta}$ island \cite{Gustafsson}.

In our previous paper \cite{NagaiNano}, we have shown, by solving the Bogoliubov-de Gennes (BdG) equations
self-consistently for the $d$-wave order parameter in diamond-shaped nanoislands with [110] surfaces, that spontaneous TRS breaking occurs at
low temperatures via the spontaneous emergence of a new complex order parameter. 
The spontaneous disappearance of the topological protection is accompanied by this order parameter, induced on the surfaces below 
a critical temperature lower than the superconducting transition temperature, $T_{\rm c}$. 
The additional order parameter has extended $s$-wave symmetry and characterizes the energy 
splitting of the Andreev bound states on the surfaces.
We have also shown that integer vortex-antivortex pairs are formed in the extended $s$-wave order parameter along the surfaces if the side length of a nanoisland is relatively large \cite{NagaiNano}.

In this paper, we consider nanoribbons made of a $d$-wave superconductor by self-consistently solving the BdG equations along with the $d$-wave gap equation. 
The periodic boundary condition allows us to model nanoribbons of infinite length.
We fix the temperature such that a TRS-breaking phase occurs.
We find a spatial distribution of vortex-antivortex pairs along the [110] surfaces, which is similar to that in a large nanoisland \cite{NagaiNano}.

\section{Model and method}
\subsection{Model}
We consider a tight-binding Hamiltonian ${\cal H} = {\cal H}_{\rm BCS} + {\cal H}_{\rm fab}$ on a two-dimensional square lattice. 
Here the generalized Bardeen-Cooper-Schrieffer (BCS) Hamiltonian for $d$-wave superconductivity is ${\cal H}_{\rm BCS} = \sum_{ij,\sigma} (-t_{ij} - \mu) c_{i \sigma}^{\dagger} c_{j \sigma} + \sum_{ij} \left[ 
\Delta_{ij} c_{i \up}^{\dagger} c_{j \down}^{\dagger} + {\rm H.c.} 
\right]$, where $c_{i \sigma}^{\dagger}$ creates the electron with spin $\sigma$ at site $i$ and $\mu$ denotes the chemical potential. 
For simplicity, we only consider the nearest-neighbor hopping.
We use the unit system with $\hbar = k_{\rm B} = 1$. 
The Hamiltonian for the fabrication potential ${\cal H}_{\rm fab}$ can be written as ${\cal H}_{\rm fab} = \sum_{i,\sigma} V_{0}(\Vec{r}_{i}) c_{i \sigma}^{\dagger} c_{i \sigma}$, 
where $\Vec{r}_{i}$ is a position vector with index $i$ whose origin is a center of the system: $\Vec{r}_{i} = (i_{x} - L_{x}/2-1/2, i_{y}-L_{y}/2-1/2)$.  
The fabrication potential term ${\cal H}_{\rm fab}$ enables us to create an arbitrary shape for the nanoscale superconductor.
One diagonalizes ${\cal H}$ to solve the BdG equations expressed as 
\begin{align}
\sum_{j} \hat{H}_{ij}
\left(\begin{array}{c}
u_{\gamma}(\Vec{r}_{j}) \\
v_{\gamma}(\Vec{r}_{j}) 
\end{array}\right)
&= 
E_{\gamma}
\left(\begin{array}{c}
u_{\gamma}(\Vec{r}_{i}) \\
v_{\gamma}(\Vec{r}_{i}) 
\end{array}\right), \label{eq:bdg}
\end{align}
with
\begin{align}
\hat{H}_{ij} &= 
 \left(\begin{array}{cc}
\left[ \hat{H}^{\rm N} \right]_{ij} & [\hat{\Delta}]_{ij} \\
\left[ \hat{\Delta}^{\dagger} \right]_{ij}  & -\left[ \hat{H}^{\rm N \ast} \right]_{ij}
\end{array}\right).
\end{align}
Here $\left[ \hat{H}^{\rm N} \right]_{ij} = - t_{ij} - (\mu - V_{0}(\Vec{r}_{i}) ) \delta_{ij}$ 
and $[\hat{\Delta}]_{ij} = V_{ij} \sum_{\gamma=1}^{2N} u_{\gamma}(\Vec{r}_{j}) v_{\gamma}^{\ast}(\Vec{r}_{i}) f(E_{\gamma})$, 
where $N$ is the number of lattice sites, $V_{ij}$ denotes the pairing interaction, and $f(x)$ is the Fermi-Dirac distribution function. 
The $d$-wave order parameter at site $i$ is defined in terms of the mean fields as
\begin{equation}
\Delta_{d,i} =  (\Delta_{\hat{x},i}+\Delta_{-\hat{x},i} - \Delta_{\hat{y},i} - \Delta_{-\hat{y},i})/4, \label{eq:dwave}
\end{equation}
with $\Delta_{\pm \hat{e},i} = \Delta(\Vec{r}_{i},\Vec{r}_{i} \pm \hat{e})$, where $\hat{x}$ and $\hat{y}$ denote the unit vectors in the square lattice. 
Considering two-dimensional systems, we assume that the field penetration depth $\lambda$ is infinity.
We use complex mean fields $\{\Delta_{ij}\}$
with random phases as an initial guess for our calculation.
For the results presented below, the chemical potential is fixed to be $\mu = -1.5t$ and the pairing interaction is nonzero only between nearest-neighbor sites, 
$V_{ij} \equiv U = -2t$, so that the resulting order parameter is purely of the $d_{x^{2}-y^{2}}$-wave symmetry in the TRS-preserved phase.

\begin{figure}[t]
\begin{center}
     \begin{tabular}{p{ 0.3 \columnwidth}} 
      \resizebox{0.3 \columnwidth}{!}{\includegraphics{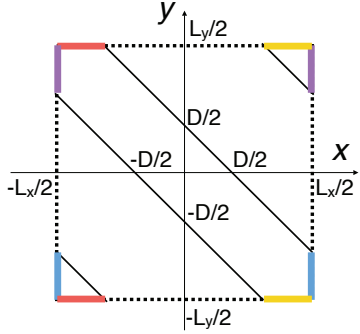}} 
    \end{tabular}
\end{center}
\caption{
(Color online) Schematic figure of a nanoribbon superconductor. The periodic boundary condition is used. 
\label{fig:fig1}
 }
\end{figure}

In a diamond-shaped square nanoisland of a $d$-wave superconductor, 
we find at a certain temperature that 
the system goes through a second-order phase transition into a ground state with the property that the 
extended $s$-wave order parameter defined by \cite{Kosztin,Hosseini} 
\begin{align}
\Delta_{s,i} =  (\Delta_{\hat{x},i}+\Delta_{-\hat{x},i} + \Delta_{\hat{y},i} + \Delta_{-\hat{y},i})/4, 
\end{align}
becomes nonzero along the [110] surfaces \cite{NagaiNano}. 
We stress that no on-site attractive interaction is involved in our calculation: a $d+is$ state typically appears in the presence of such an on-site interaction \cite{Kashiwaya,Black-Schaffer}. 
The fabrication potential of a nanoribbon is 
\begin{align}
V_{0}(\Vec{r}_{i}) &= \begin{cases}
0   & -x_{i} - \frac{D}{2} < y_{i} < - x_{i} + \frac{D}{2},  \\
0   & -x_{i} - \left( \frac{L_{x} + L_{y} - D}{2} \right) < y_{i} < - x_{i} + \left( \frac{L_{x} + L_{y} - D}{2} \right),  \\
500t & {\rm else},
\end{cases}
\end{align}
where the nanoribbon width is $D/\sqrt{2}$. 
Our previous study \cite{NagaiNano} indicates the presence of a suitable spatial scale for the formation of vortex-antivortex pairs; the width of a nanoribbon being roughly eight times the coherence length. 
Figure~\ref{fig:fig1} shows our setup of a $d$-wave superconducting nanoribbon, with the periodic boundary condition.

\subsection{Method}
To solve the BdG equations and the gap equation self-consistently, 
we calculate the mean fields, $[\hat{\Delta}]_{ij} = \langle c_{i} c_{j} \rangle$. 
With the use of analytic continuation, the mean fields can be rewritten as 
\begin{align}
[\hat{\Delta}]_{ij} &= T \sum_{n=-\infty}^{\infty} [\check{G}(i \omega_{n})]_{ji+N}, \label{eq:dij}
\end{align}
with the fermion Matsubara frequency $\omega_{n} = \pi T (2 n + 1)$. 
Here, $\check{G}(i \omega_{n})$ is 
a $2N \times 2N$ matrix form of the temperature-dependent Green's function, 
\begin{align}
\check{G}(i \omega_{n}) &= \left[i \omega_{n} \check{1} - \check{H} \right]^{-1}.
\end{align}
Since the element is calculated as 
$\left[ \check{G}(i \omega_{n}) \right]_{ij} = \Vec{e}^{i} \check{G}(i \omega_{n}) \Vec{e}^{j}$, 
we obtain
\begin{align}
\left[ \check{G}(i \omega_{n}) \right]_{ij} &= \Vec{e}^{i} \Vec{x}^{j}(n), 
\end{align}
with the solutions of the linear equations, 
\begin{align}
\left[i \omega_{n} \check{1} - \check{H} \right]  \Vec{x}^{j}(n) &= \Vec{e}^{j},
\end{align}
where $[\Vec{e}^{i}]_{k} = \delta_{ik}$.
These linear equations for different frequencies 
can be solved simultaneously by the reduced-shifted-conjugate-gradient (RSCG) method \cite{NagaiRSCG}.
The cutoff for the Matsubara frequency is $\omega_{\rm c} = 60\pi = \pi T(2 n_{c}+1)$. 
The convergence criterion for the RSCG method is that the residual is smaller than 0.1, leading to a sufficient accuracy in the calculation of the Green's function \cite{NagaiRSCG}.

To find the true ground state, we calculate the thermodynamic potential given by \cite{NagaiNano}, 
\begin{align}
\Omega_{s} &= - T \sum_{\gamma=1}^{2N} \ln \left[
1 + \exp \left( \frac{E_{\gamma}}{T} \right)
 \right] - \sum_{ij} \frac{|[\hat{\Delta}]_{ij} |^{2}}{U}.
\end{align}
Once the mean fields are converged in the RSCG method, all the eigenvalues, 
$E_{\gamma}$, are obtained by the Sakurai-Sugiura (SS) method \cite{Sakurai,NagaiSS}.
The SS method allows us to extract the eigenvalues (and corresponding eigenvectors) of a generic matrix from a given domain on the complex plane \cite{Sakurai}.
We take the adequate number of small circular domains for calculating all of the eigenvalues.

\subsection{Topological invariant}
A topological invariant characterizing $d$-wave superconductors with TRS is the one-dimensional (1D) topological invariant \cite{Satoflat}, 
defined by the accumulation of the phase of the quasiparticle wavefunction along a closed loop in the momentum space (i.e., winding number), 
\begin{align}
w(k_{y}^{0}) &= \frac{1}{2}
{\rm sgn} \: \left[ 
\frac{\partial \epsilon(-k_{x}^{0},k_{y}^{0})}{\partial k_{x}} \right] 
\left(
{\rm sgn} \: \left[ \Delta(-k_{x}^{0},k_{y}^{0}) \right]-{\rm sgn} \: \left[ \Delta(k_{x}^{0},k_{y}^{0}) \right]
\right),
\end{align}
where $\epsilon(\Vec{k})$ is the normal-state band dispersion and $(\pm k_{x}^{0},k_{y}^{0})$ denote the intersection points between the integration path with fixed $k_{y}^{0}$ and the Fermi surface. 
Thus, if the gap function $\Delta(\Vec{k})$ satisfies 
\begin{align}
\Delta(-k_{x}^{0},k_{y}^{0}) \Delta(k_{x}^{0},k_{y}^{0}) < 0, \label{eq:cond}
\end{align}
then the 1D topological invariant has a nonzero value and the zero-energy bound states 
exist because of the bulk-edge correspondence. 
Varying $k_{y}^{0}$ in the range with no gap closing, we can obtain the surface 
dispersion relation for the zero-energy bound states~\cite{Satoflat}. 
Since the time-reversal operation includes the complex conjugation operator, 
the gap function $\Delta$ should be real in the TRS-preserved phase. 
In contrast, if the gap function is complex (TRS-breaking phase), the above winding number becomes ill-defined \cite{NagaiNano}. 

\begin{figure}[t]
\begin{center}
     \begin{tabular}{p{ 1 \columnwidth}} 
      \resizebox{1\columnwidth}{!}{\includegraphics{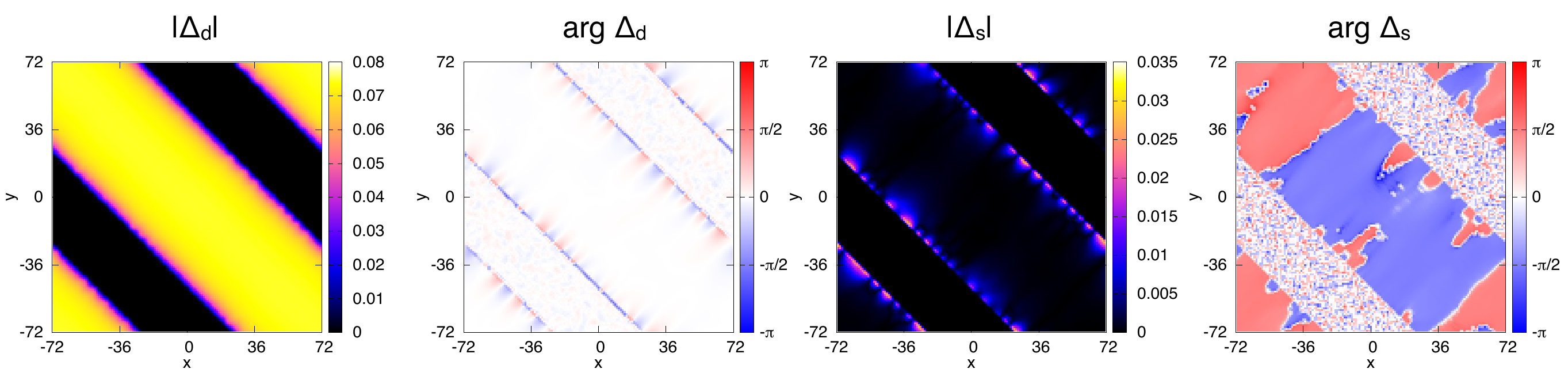}} 
    \end{tabular}
\end{center}
\caption{
(Color online) $d$-wave and induced extended $s$-wave order parameters in the TRS-broken phase. The system size is $L_{x} \times L_{y} = 144 \times 144$ and the width of the nanoribbon is $D/\sqrt{2} = (2L_{x}/3)/\sqrt{2}$. The temperature is $T = 0.01t$. 
\label{fig:fig2}
 }
\end{figure}

\begin{figure}[t]
\begin{center}
     \begin{tabular}{p{ 1 \columnwidth}} 
      \resizebox{1\columnwidth}{!}{\includegraphics{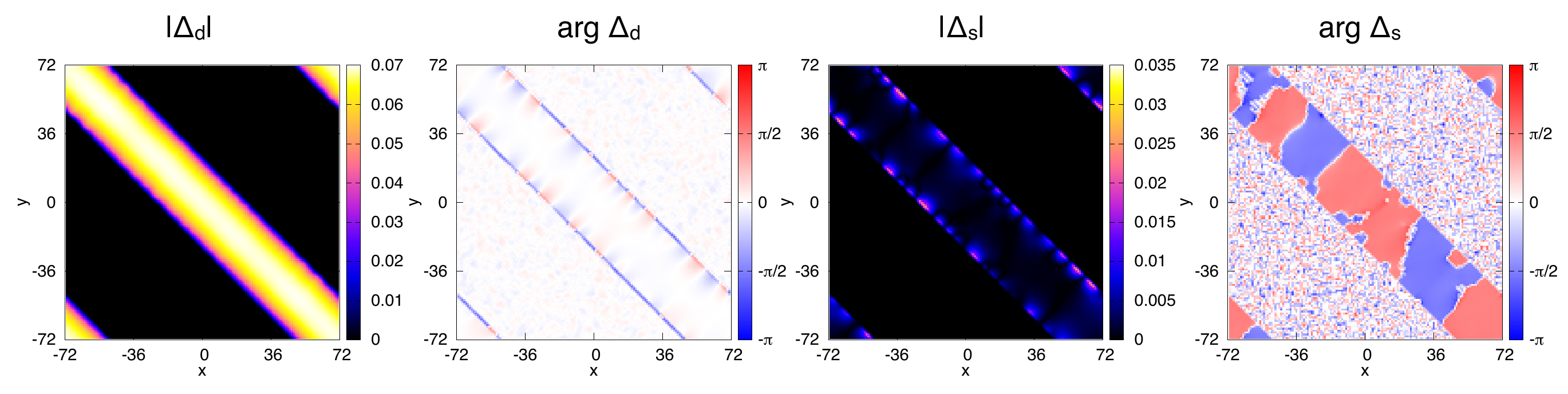}} 
    \end{tabular}
\end{center}
\caption{
(Color online) $d$-wave and induced extended $s$-wave order parameters in the TRS-broken phase. The system size is $L_{x} \times L_{y} = 144 \times 144$ and the width of the nanoribbon is $D/\sqrt{2} = (L_{x}/3)/\sqrt{2}$. The temperature is $T = 0.01t$. 
\label{fig:fig3}
 }
\end{figure}

\begin{figure}[t]
\begin{center}
     \begin{tabular}{p{ 0.5 \columnwidth}p{ 0.5 \columnwidth}} 
      \resizebox{0.5\columnwidth}{!}{\includegraphics{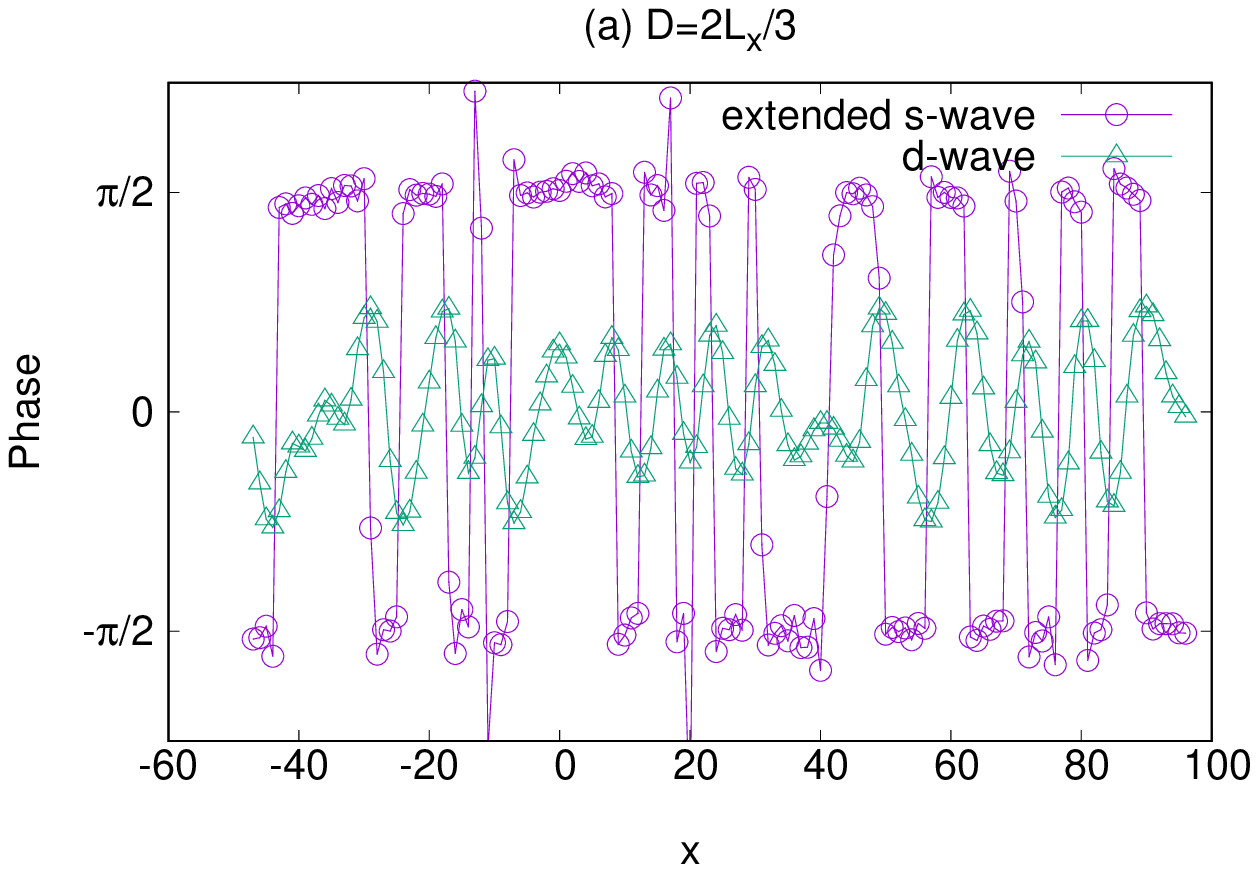}} 
      &
       \resizebox{0.5\columnwidth}{!}{\includegraphics{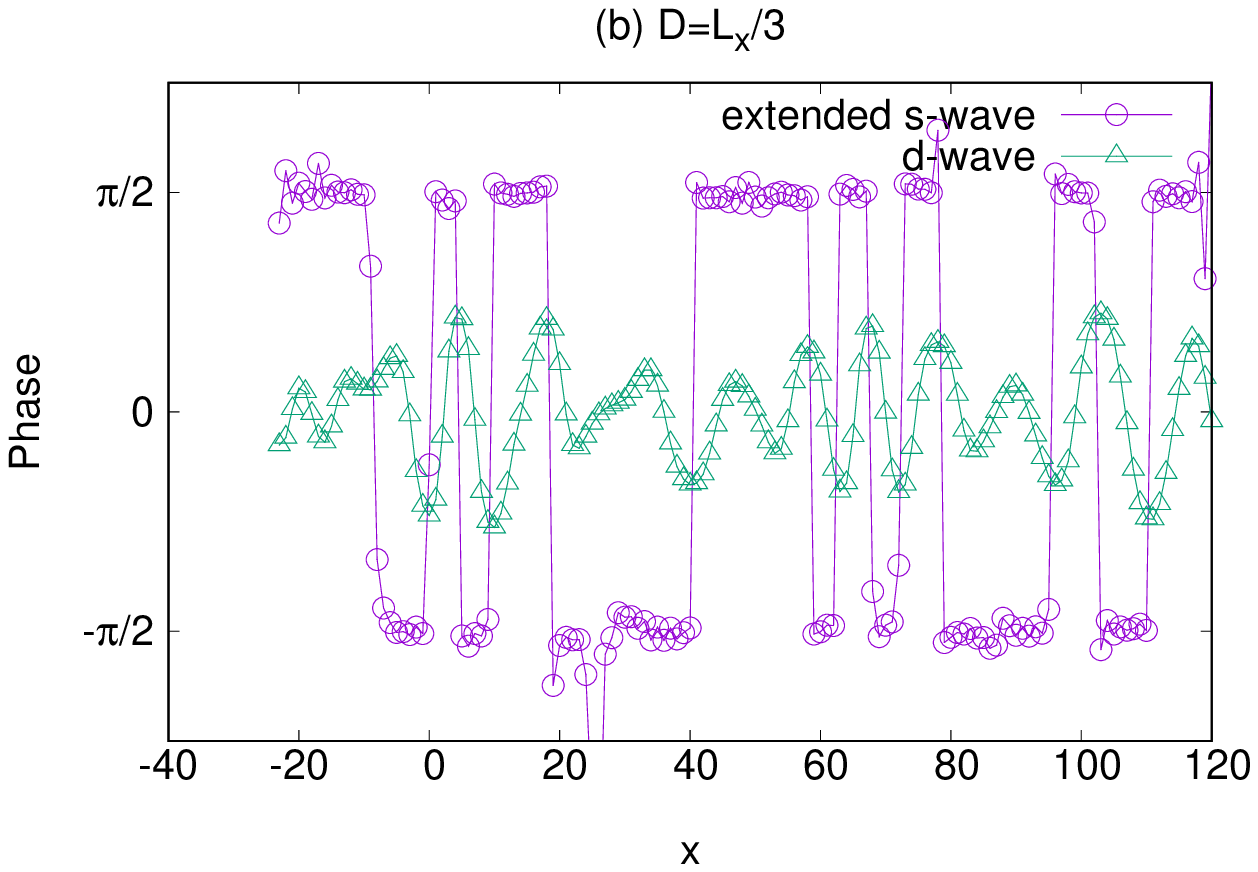}} 
    \end{tabular}
\end{center}
\caption{
(Color online) The spatial distributions of the phases 
${\rm arg} (\Delta_{s})$ and ${\rm arg} (\Delta_{d})$ 
on the surface edge ($y = -x + D/2$).
\label{fig:fig4}
 }
\end{figure}

\section{Results}
We consider two kinds of 
$d$-wave superconducting nanoribbons, with width $D =2L_{x}/3$ and $D = L_{x}/3$. 
Both of these systems have $L_{x} \times L_{y} = 144 \times 144$ lattice sites. 
We set the temperature $T = 0.01t$, where the 
TRS-breaking phase appears in a diamond-shape nanoisland \cite{NagaiNano}. 
Figures~\ref{fig:fig2} ($D=2L_x/3$) and \ref{fig:fig3} ($D=L_x/3$) show the amplitudes and phases of the order parameters. 
Focusing on the spatial distribution of arg($\Delta_d$) and arg($\Delta_s$), 
we find that in the extended $s$-wave order parameter the phase varies from $\pi/2$ to $-\pi/2$, and vice versa, with an alternating manner. 
Thus, the $d$-wave and extended $s$-wave order parameters are both complex and TRS is broken 
at this temperature. 
The ground state in this phase is fundamentally different from a $d+is$ state, which was discussed previously \cite{Matsumoto,Fogelstroem,Black-Schaffer}.
We also confirm that there is no zero-energy surface state in this phase.

Let us now focus on the spatial distribution of the order parameter phases in Figs.~\ref{fig:fig2} and \ref{fig:fig3}. 
To this end, we examine the phase distribution on the surface, $y = -x + D/2$, 
as seen in Fig.~\ref{fig:fig4}. 
It can be seen more clearly that arg($\Delta_s$) spatially varies with $\pi$ shifts 
along the surfaces.
In contrast, ${\rm arg}\:(\Delta_d)$ has no such phase shift and changes very little, although it is important to note that the phase is nonzero and hence $\Delta_d$ itself is complex.
The amplitude of the extended $s$-wave order parameter gives us insight into this phenomenon: 
it vanishes along the surface as seen in Figs.~\ref{fig:fig2} and \ref{fig:fig3}, at positions where the $\pm \pi$ jump occurs in the phase.
Thus, integer vortex-antivortex pairs appear in the extended $s$-wave order parameter along the surfaces.
This result is similar to that for a larger-size $d$-wave superconducting nanoisland \cite{NagaiNano}. 
We note that the positions of the vortex-antivortex pairs depend on the initial condition of the randomly-distributed phase of the $d$-wave order parameter. 
Such vortex-antivortex pairs can be detected by surface-sensitive probes. 

Vorontsov has proposed that films of a $d$-wave superconductor at low temperatures can exhibit unusual superconducting phases due to transverse confinement, associated with the spontaneous breaking of either TRS or continuous translational symmetry \cite{Vorontsov}.
In the current work,
the ground state still has the translational symmetry in the bulk region of the nanoribbons. 
According to Vorontsov's proposal, which is based on the quasiclassical Eilenberger approach, nanoribbons much narrower than the present ones can have a translational-symmetry breaking phase. 
The coherence length in our calculations would be too small to study such extremely narrow nanoribbons. 
An interesting future work is to examine such breaking of translational symmetry within the BdG framework by decreasing the width of nanoribbons.

In conclusion, we have shown by solving the BdG equations self-consistently for $d$-wave nanoribbons that spontaneous emergence of a complex, extended $s$-wave order parameter along the [110] surfaces breaks time-reversal symmetry. 
Such TRS-breaking phases can also sustain vortex-antivortex pairs along the surfaces.

\ack
The calculations were performed on the supercomputing system SGI ICE X at the Japan Atomic Energy Agency. 
This study was partially supported by the ``Topological Materials Science'' (No. JP16H00995) KAKENHI on Innovative Areas from JSPS of Japan, and the Natural Sciences and Engineering Research Council of Canada.

\end{document}